\documentstyle[preprint,aps]{revtex}
\begin{document}
\title{Two-dimensional Yang--Mills theory in the leading 1/N expansion 
revisited}
\author{A. Bassetto ($^*$)}
\address{CERN, Theory Division, CH-1211 Geneva 23, Switzerland\\
INFN, Sezione di Padova, Padua, Italy}
\author{G. Nardelli}
\address{Dipartimento di Fisica, Universit\`a di Trento,  
38050 Povo (Trento), Italy \\ INFN, Gruppo Collegato di Trento, Italy}
\author{A. Shuvaev}
\address{Theory Department, St.Petersburg Nuclear Physics Institute\\
188350, Gatchina, St.Petersburg, Russia}
\maketitle
\begin{abstract}
We obtain a $formal$ solution of an integral equation for $q\bar q$ bound
states, depending on a parameter $\eta$ which interpolates between 't Hooft's
($\eta=0$) and Wu's ($\eta=1$) equations. We also get an explicit 
approximate expression for its spectrum for a particular value of the ratio
of the coupling constant to the quark mass.
The spectrum turns out to be in qualitative
agreement with 't Hooft's as long as $\eta \neq 1$. In the limit
$\eta=1$ (Wu's case) the entire spectrum collapses to zero, in particular 
no rising Regge trajectories
are found.
\end{abstract} 

\noindent
($^*$) On leave of absence from
Dipartimento di Fisica, Via Marzolo 8 -- I-35131
Padova.
\vfill\eject

\narrowtext

\section{The $q \bar q$ bound state equation and its formal solution}
\noindent
In 1974 G. 't Hooft \cite{TH74} proposed a very interesting model to describe
mesons, starting from an SU(N) Yang--Mills theory in 1+1 dimensions
in the large-$N$ limit.

Quite remarkably in this model quarks look confined, while a discrete
set of quark-antiquark bound states emerges, with squared masses lying
on rising Regge trajectories.

The model is solvable thanks to the ``instantaneous'' character of
the potential acting between quark and antiquark.

After that pioneering investigation, many interesting papers
followed 't Hooft's approach, pointing out further remarkable
properties of his theory and blooming into the recent achievements
of two-dimensional QCD \cite{WTT1},\cite{WTT2},\cite{BT92}.

Three years later such an approach was criticized by T.T. Wu \cite{WU77},
who replaced the instantaneous 't Hooft potential by an expression
with milder analytical properties, allowing for a Wick rotation
without extra terms.
Actually this expression is nothing but the (1+1)-dimensional 
version of the Mandelstam-Leibbrandt (ML)\cite{ML84}
propagator, a choice which is mandatory
in order to achieve gauge
invariance and renormalization in 1+3 dimensions \cite{BNS91},\cite{BKKN},
\cite{BDG94}.

Unfortunately this modified formulation led to a quite involved bound
state equation. An attempt to treat it
numerically in the zero bare mass case for quarks \cite{BS78} led only to
partial answers in the form of a completely different physical
scenario. In particular no rising Regge trajectories were found.

The integral equation for the quark self-energy in the Minkowski
momentum space is

\begin{eqnarray}
\label{unob}
\Sigma(p;\eta)&=& i {{g^2}\over {\pi^2}} {{\partial}\over {\partial p_{-}}}
\int dk_{+}dk_{-} \Big[P\Big({{1}\over {k_{-}-p_{-}}}\Big)+
i \eta \pi \ {\rm sign} (k_{+}-p_{+}) \delta (k_{-}-p_{-})\Big]\nonumber\\
&\cdot&{{k_{-}}\over {k^2+m^2-k_{-}\Sigma (k;\eta)-i\epsilon}},
\end{eqnarray}
where $P$ denotes the Cauchy principal value prescription
(CPV), $g^2=g_0^2 \,N$, and $\eta$ is a parameter that is used
to interpolate between 't Hooft's ($\eta =0$) and Wu's equation ($\eta =1$).

Its exact solution reads

\begin{eqnarray}
\label{uno}
\Sigma(p;\eta)&=& {{1}\over {2p_{-}}}\left(\Big[p^2+m^2+(1-\eta){{g^2}\over
{\pi}}\Big]-\right.\nonumber\\
&-&\left.\sqrt {\Big[p^2+m^2-(1-\eta){{g^2}\over
{\pi}}\Big]^2 - {4\eta g^2 p^2\over \pi}}\,\,\right)\ ,
\end{eqnarray}
where the boundary condition has been chosen in such a way that 
$p_{-}\Sigma(p^2=+\infty)=g^2/\pi$. When continuing in $p^2$, care
is to be taken in the choice of the square-root determination.

One can immediately realize that 't Hooft's and Wu's solutions
are recovered for $\eta =0$ and $\eta =1$, respectively.

The dressed quark propagator turns out to be

\begin{equation}
\label{due}
S(p;\eta) = - {{i p_{-}}\over {m^2+2 p_{+}p_{-}- p_{-}\Sigma(p;\eta)}}
\end{equation}
and the equation for a $q \bar q$ bound state in
Minkowski space, using light-cone coordinates, is

\begin{eqnarray}
\label{tre}
\psi(p,r)&=& {{-ig^2}\over {\pi ^2}} S(p;\eta) S(p-r;\eta)
\int dk_{+}dk_{-} \Big[P\Big({{1}\over
{(k_{-}-p_{-})^2}}\Big)-\nonumber\\
&-&
i \eta \pi \,sign (k_{+}-p_{+}) \delta^{\prime} (k_{-}-p_{-})\Big]
\psi(k,r).
\end{eqnarray}
We are here considering for simplicity the equal mass case.

The 't Hooft potential ($\eta=0$) exhibits an infrared singularity,
which was handled, in the original formulation, by introducing
an infrared cutoff; a quite remarkable feature of this theory
is that bound state wave functions and related eigenvalues
turn out to be cutoff-independent. Actually, in
ref. \cite{CA76}, it has been pointed out that the singularity
at $k_{-}=0$ can also be regularized by the
CPV without altering gauge
invariant quantities. Then, the difference between the two
cases $\eta=1$ and $\eta=0$ 
is represented by the following distribution

\begin{equation}
\label{unoa}
\Delta (k)\equiv {{1}\over {(k_{-}-i\epsilon
\,\,{\rm sign} (k_{+}))^2}} - P\Big({{1}\over {k_{-}^2}}\Big)= - i \pi
\,\,{\rm sign} (k_{+}) \delta^{\prime}(k_{-}).
\end{equation}
both in the equation for the self-energy and in the one for $q\bar q$ bound
states.

In ref.\cite{BG96}, 
it has been shown that, starting from 't Hooft's solutions, 
no correction affects 't Hooft's spectrum
when the difference in eq.\,(\ref{unoa}) is treated as a single insertion
both in the ``potential''  and in the propagators.  
Wu's equation for colourless bound states,
although much more involved than the
corresponding 't Hooft one, might still apply, according to the
heuristic lesson one learns from a single insertion.

\smallskip

It is the purpose of this paper to show that
unfortunately this conclusion is unlikely to persist beyond a single
insertion. This should not come as a surprise, since
Wu's equation is deeply
different from 't Hooft's and might be related to a different 
physical scenario (see for instance
\cite{ZY95}).

\smallskip

It is useful to
introduce dimensionless variables $x$, $y$ and
$\alpha$ \begin{eqnarray} p_-\,&=&\,x\,r_-
\nonumber \\ p_+\,&=&\,y\,r_+ \nonumber \\
2\,r_+r_-\,&=&\,-\alpha m^2, \nonumber \end{eqnarray}
so that $$
\frac{p^2}{m^2}\,=\,\frac{2\,p_+p_-}{m^2}\,=
\,xy\,\frac{2\,r_+r_-}{m^2}\,=
\,-\alpha\,xy.$$ 

\smallskip

In these notations the quark
propagators are $$
S(p)\,=\,-\frac{ir_-}{m^2}\,\frac x {1-\alpha xy
- \Sigma(\alpha xy)}
\,=\,-\frac{ir_-}{m^2}\,S(x,y), $$ $$
S(p-r)\,=\,\frac{ir_-}{m^2}\,\frac {\overline x}
{1-\alpha\overline  x \overline y - \Sigma(\alpha
\overline x \overline y)}
\,=\,\frac{ir_-}{m^2}\,S(\overline x,\overline y),
$$ where $$ \overline x \equiv 1-x, \quad
\overline y \equiv 1-y, $$ \begin{eqnarray}
\Sigma\,&=&\,\frac 12 \left(\bigl[1-\alpha xy +
(1-\eta)\frac{g^2}{\pi m^2} \bigr]\,+\right.
 \nonumber \\ &+&\left.  \sqrt{
\,[1-\alpha
xy - (1-\eta)\frac{g^2}{\pi m^2}\bigr]^2\,+4\eta
\frac{g^2 \alpha xy}{\pi m^2}\,} \,\,\right).\nonumber
\end{eqnarray} 

Notice that the square root has changed sign as it has been continued
to positive values of $\alpha$.

\smallskip

The bound state equation takes the
form 
\begin{eqnarray} 
\label{ttwu}
\psi(x,y)\,&=&\,-\frac{ig^2}{2\pi^2 m^2}\,\alpha\,
S(x,y)S(\overline x, \overline y)\,\int dx^\prime
dy^\prime\, \Bigl[\,P\frac 1{(x^\prime -x)^2}\,+
\\ &+&\,i \eta \pi \,{\rm sign}(y^\prime - y)\,
\delta^\prime (x^\prime - x)\Bigr]\,
\psi(x^\prime, y^\prime).\nonumber \end{eqnarray}

\vskip .5truecm

We write eq.\,(\ref{ttwu}) symbolically as

\begin{equation}
\label{symb}
\psi(x,y)\,=\,{\cal S}\,\int_{-\infty}^\infty dy^{\prime}
\bigl[\{H\,\psi\}(x,y^{\prime})\,-i\,\eta \pi\, {\rm sign}(y^{\prime}-y)\,
\partial_{x}
\psi(x,y^{\prime})\bigr]. 
\end{equation}

Here $H$ denotes 't Hooft's operator
$$\{H\,\psi\}(x,y)\,=\,\int_{-\infty}^\infty dx^{\prime}
\,P\frac 1{(x^{\prime}- x)^2} \psi(x^{\prime},y) $$ 
and ${\cal S}=-\frac{ig^2}{2 \pi^2 m^2}\alpha  S(x,y )S(\bar x, \bar y)$
is a multiplication operator.

\smallskip

After introducing the function $$
F(x,y)\,=\,\int_{-\infty}^y dy^{\prime}\,\psi(x,y^{\prime}), \qquad
F(x,-\infty)\,=\,0 $$ the equation takes the
form 
\begin{equation}
\label{form}
\frac{\partial F}{\partial y}\,=\,{\cal S}\,\{\bigl(\,H\,-\,i\pi
\eta\,\partial_{x}\,\bigr)F\}(x,\infty)\,+\, 2\pi i
\eta\,{\cal S}\,\partial_{x}\,F(x,y).  
\end{equation}

The {\it formal} solution
of this equation is \begin{eqnarray}\label{formal}
F(x,y)\,&=&\,P\exp\left\{2\pi i \eta\int_{-\infty}^y
dy^\prime \,{\cal S}(x,y^\prime)\,\partial_{x} \right\}\,\cdot
\nonumber \\ \cdot\,\int_{-\infty}^y
&dw&\,\left[P\exp\left\{2\pi
i \eta\int_{-\infty}^w dy^{\prime} 
\,{\cal S}(x,y^\prime)\,\partial_{x}\right\}\,\right]^{-1}\cdot\nonumber\\
&\cdot&{\cal S}(x,w)\{\bigl(H - i\pi\eta \partial_{x}\bigr)F\}(x,\infty),
\end{eqnarray} 
where the ``path-ordered"
exponent appears since the operators ${\cal S}$
and $\partial_{x}$ do not commute. 

For $y=+\infty$, one gets
the closed one-dimensional equation for the
function $F(x,\infty)$ (which is just the analogue of 't Hooft's
wave function):  
\begin{eqnarray}
\label{Hoof}
F(x,\infty)\,&=&\,\int_{-\infty}^\infty
dy\,P\exp\left\{2\pi i\eta\int_y^{\infty}
dy^\prime \,{\cal S}(x,y^\prime)\,\partial_{x}\right\}\cdot\nonumber\\
&\cdot&{\cal S}(x,y)\{\bigl(H - i\pi\eta \partial_{x}\bigr)F\}(x,\infty).
\end{eqnarray}

An equivalent
form is 
\begin{eqnarray}
\label{equi}
F(x,\infty)\,&=&\,i\,\int_{-\infty}^\infty dy\,\frac
1{2\pi\eta} \frac \partial {\partial
y}\,P\exp\left\{2\pi i \eta\int_y^{\infty}
dy^\prime
\,{\cal S}(x,y^\prime)\,\partial_{x}\right\}\,\cdot\nonumber \\
&\cdot& \frac 1{\partial_{x}} \,\{\bigl(H -i \pi\eta
\partial_{x}\bigr)F\}(x,\infty),
\end{eqnarray} 
or,
finally, 
\begin{eqnarray} \label{Finf}
F(x,\infty)\,&=&\,\frac
i{2\pi\eta}\left[I\,-\,P\exp\left\{2\pi i \eta
\int_{-\infty}^{\infty}dy^\prime
\,{\cal S}(x,y^\prime)\,\partial_{x}\right\}\right]\, \cdot\nonumber \\
&\cdot&\frac 1{\partial_{x}}\, \{\bigl(H -i \pi\eta
\partial_{x}\bigr)F\}(x,\infty).
\end{eqnarray}

We notice that, in the limit $\eta =0$, 't Hooft's equation is
correctly reproduced.
Once eq.\,(\ref{Finf}) is solved, eq.\,(\ref{formal})
provides us with the solution of the original two-variable equation
(\ref{ttwu}).

\section{An analytical approximate solution}
\noindent
Equations (\ref{Finf})and (\ref{ttwu}) are by far too difficult to be
concretely solved by means of analytical procedures (and probably
also numerically). There is, however, a particular case in which
they can be successfully tackled, at least heuristically.
This case is realized when $\Sigma(\alpha xy)\simeq 1$, which in turn
means a peculiar value of the
ratio ${g^2\over {\pi m^2}} \simeq 1$. 

\smallskip

In such a case the operator ${\cal S}$ takes the form
\begin{equation}
\label{dueuno}
{\cal S}(x,y)= -\,{{ig^2}\over{2\pi ^2 \alpha m^2}}
\qquad{1\over{y+i\epsilon \,\, {\rm sign}(x)}}\qquad
{1\over{\bar y+i\epsilon \,\, {\rm sign}(\bar x)}}.
\end{equation}

From here on we shall rely on this particular form of ${\cal S}$.
Such a form can also be approximately obtained
for a generic value of the coupling in the limit of large values
of $\alpha$.

In future developments the following integral will be needed:
$$\int_{-\infty}^{+\infty} dy {\cal S}(x,y)=-\lambda \theta(x)
\theta(\bar x),$$ $\theta$ being the usual step distribution and
$\lambda= {{g^2}\over {\alpha \pi m^2}}$.

The commutator between ${\cal S}$ and $\partial_{x}$ is a measure
concentrated at the values $x=0$ and $x=1$. In the Appendix
we shall argue that, whenever the product $\eta \lambda$ is not
pure imaginary, the contribution
from that commutator vanishes:
``path-ordered'' exponentials can be turned into ``normal-ordered'' ones.

\smallskip
Then eq.\,(\ref{Finf}) can be written as

\begin{eqnarray} 
\label{duedue}
F(x,\infty)\,&=&\,\frac
i{2\pi\eta}\left[I\,-\,N\exp\left\{-2\pi i \eta \lambda \theta (x)
\theta (\bar x)
\,\partial_{x}\right\}\right]\, \cdot\nonumber \\
&\cdot&\frac 1{\partial_{x}}\, \{\bigl(H -i \pi\eta
\partial_{x}\bigr)F\}(x,\infty).
\end{eqnarray}

This equation can be approximately diagonalized by a Fourier
transform

\begin{equation}
\label{duetre}
\tilde F(k)=\,\int_{-\infty}^{+\infty}\,e^{ikx}F(x,\infty)\,dx.
\end{equation}

Equation (\ref{duedue}) indeed becomes

\begin{equation}
\label{duetrebis}
\tilde F(k)={1 \over{4\pi 
\eta}}\int_{-\infty}^{+\infty}dq\, \tilde F(q)e^{i{{k-q}\over 2}}\, 
{{\sin{{k-q}\over 2}}
\over {{k-q}\over 2}}
\,({\rm sign} \,(q) +\eta) \,(1-e^{-2\pi q\eta \lambda}), 
\end{equation}
where the well-known relation $$\exp\{\Lambda \partial_{x}\}f(x)= f(x+
\Lambda)$$ has been used. 

Taking the approximation ${\sin x \over x}\simeq \pi\,\delta (x)$ into
account, we get

\begin{equation}
\label{duequattro}
\tilde F(k) \cdot \Big[({\rm sign}(k)+\eta) \exp(-2 \pi \eta\lambda k)- 
({\rm sign}(k)-\eta)
\Big]\simeq 0.
\end{equation}

In order to get non-vanishing solutions, the equation

\begin{equation}
\label{duecinque}
\lambda |k| = {1\over {2\pi \eta}} {\rm Log} {{1+\eta}\over {1-\eta}}
\end{equation}
has to be satisfied. Here the Log has to be interpreted 
as a multivalued function.

The condition $F(0,\infty)=0$ entails the choice

\begin{equation}
\label{extra}
\tilde F(k)= C(k)\cdot \delta[|k|-{1\over{2\pi \eta \lambda}}{\rm Log}{{1+\eta}
\over {1-\eta}}],
\end{equation}
where $C(k)$ is an odd function of $k$.

Finally the condition that $F(1,\infty)=0$ 
induces the eigenvalues

\begin{equation}
\label{duesei}
|k_n|= \,n\pi,\qquad\qquad\qquad\qquad n>0,
\end{equation}
namely

\begin{equation}
\label{duesette}
\alpha_n=\, n\,{{2\pi \eta g^2}\over{m^2 \,{\rm Log} {{1+\eta}\over{1-\eta}}}}.
\end{equation}

Each eigenvalue is thereby infinitely degenerate; but only the
principal determination of the Log reproduces 't Hooft's
spectrum in the limit $\eta =0$. The vanishing of the first-order
corrections in $\eta$ found in ref.\cite{BG96} is also confirmed.

Equation (\ref{duesette}) nicely exhibits the interpolating role of the
parameter $\eta$; eigenvalues stay discrete until the value $\eta =1$
is reached; at that value, which is by the way the value pertinent
to Wu's equation, the entire spectrum collapses to zero.

One should remember
that expression (\ref{dueuno}) is exact for the particular value
of the ratio ${{g^2}\over {\pi m^2}}= 1$. For such a tuning of
the coupling constant to the quark mass, the eigenvalues

\begin{equation}
\label{dueotto}
\alpha_n=\,n\,{{2\pi ^2  \eta}\over{{\rm Log} {{1+\eta}\over{1-\eta}}}},
\end{equation}
give the (approximate) spectrum of the theory at the ``critical''
coupling $g^2=\pi m^2$. This spectrum collapses to zero at
$\eta =1$.

The problem of studying the behaviour in a full neighbourhood of
$\eta=1$ is a difficult one. It might be that
different limits are related to possible different phases
of the theory. This interesting issue will be deferred to
future investigation.

\section{Conclusions}
\noindent

We have shown that Wu's equation for $q\bar q$ bound states,
although much more involved than the corresponding 't Hooft's one,
can nevertheless be explored, at least $formally$. We then 
obtained in a heuristic way approximate explicit results 
for a particular value
of the ratio ${g^2 \over \pi m^2}$.

To go beyond our treatment seems an arduous task. 
One perhaps needs 
mathematical theorems controlling the spectrum of the involved
operators. One could try numerical solutions of the
Wick-rotated equation and/or try to develop a perturbation theory
around the ``critical'' value $g^2=\pi m^2$. In so doing, great care
has to be taken when continuing the square root in the
exact self-energy expression.

\smallskip

In the light of the results we have found, it seems unlikely 
that Wu's equation describe the $q\bar q$ bound state 
spectrum. In spite of the singular character of its potential,
't Hooft's equation looks in a much better shape on this point.

The way of handling SU(N) Yang--Mills
theories in 1+1 dimensions by canonically quantizing 
them on the light-front, thereby reproducing 't Hooft's formulation,
might be acceptable. 
After all, in so doing, no contradiction
occurs with causality as, in strictly $1+1$ dimensions,
there are no propagating vector degrees of freedom.
Unitarity in turn is trivially satisfied, at least in the 
large-$N$ (i.e. planar) approximation.

However, one should bear in mind that 't Hooft's theory cannot be
considered as the limiting case in 1+1 dimensions of Yang--Mills theories
in higher dimensions. There, equal-time quantization
becomes compulsory in order to perform a consistent renormalization
procedure \cite{BNS91},\cite{BKKN}. On the other hand, 
a Wilson loop calculation
in $1+(d-1)$ dimensions, performed either in the Feynman gauge or in the
light-cone gauge quantized at equal time, leads to a result which,
in the limit $d=2$, cannot be reconciled 
with 't Hooft's  \cite{BKKN},\cite{BDG94}.

\vskip 1.0truecm
$Acknowledgement$

One of us (A.S.) acknowledges a grant from Istituto 
Nazionale di Fisica Nucleare (INFN).

\vfill
\eject
\centerline{\bf Appendix}
\noindent

The $P\exp$ in eq.\,(\ref{formal}) can be expressed through the normal
form
\begin{equation}
\label{Auno}
P\exp\left\{2\pi i \eta\int_{-\infty}^y
dy^\prime \,{\cal S}(x,y^\prime)\,\partial_{x}\right\}
=\,N_{x\partial_{x}}\bigl[\Phi(y,x,\partial_{x})\bigr],
\end{equation}
where the label $N_{x\partial_{x}}$ means that the operators
$x$ stand to the left of  $\partial_{x}$. 

Now we introduce the symbol of the operator 
$\Phi(y,x,\partial_{x})$, which we denote by $\Phi(y,q,p)$,
$q$ and $p$ being ``dual'' variables \cite{GI84}.
 
From the equation defining
the $P\exp$, it is easy to derive, in terms of the function $\Phi(y,q,p)$,
the relation

\begin{equation}
\label{Adue}
\frac{\partial \Phi}{\partial y}\,=\,2 \pi i \eta {\cal S}(q,y)
\,\frac{\partial \Phi}{\partial q}\,-\,2 \pi  \eta {\cal S}(q,y)\,p\Phi
\end{equation}
with the initial condition
$$
\Phi(y=-\infty)\,=\,1.
$$

One can search for solutions of the form
\begin{equation}
\label{sum}
\Phi(y,q)\,=\,\theta(-q)\,\Phi_{-\infty 0}\,+\,\theta(q)
\theta(1-q)\,\Phi_{01}\,+\,\theta(q-1)\,\Phi_{1 \infty}.
\end{equation}
Beginning with the second term and using the
relations
$$
\frac 1{y+i\epsilon \,\, {\rm sign}(q)}
\frac 1{1-y+i\epsilon \,\,{\rm sign}(1-q)}\,
\theta(q)\theta(1-q)\,=
$$
$$
=\,\frac 1{y+i\epsilon}\frac 1{1-y+i\epsilon}\,
\theta(q)\theta(1-q)
$$
$$
\frac \partial{\partial q}\,\theta(q)\theta(1-q)
\,=\,2\theta(q)\theta(1-q)\,\bigl[\delta(q)\,-\,
\delta(1-q)\bigr]
$$
($\theta(0)=1/2$ is supposed here) one gets the
equation for $\Phi_{01}$:
\begin{equation}
\label{Atre}
\frac{\partial\Phi_{01}}{\partial y}\,-\,
\varphi(y)\,\frac 1i\,\frac{\partial \Phi_{01}}
{\partial q}\,=\,\varphi(y)\bigl[p\,-\,2i\bigl(
\delta(q)-\delta(1-q)\bigr)\bigr]\,\Phi_{01},
\end{equation}

with
$$\varphi(y)\,=\,a\,\frac 1{y+i\epsilon}
\frac 1{1-y+i\epsilon},\qquad\qquad a=i \eta \lambda.
$$
After introducing the variable
$$
\xi(y)\,=\,a\bigl[\log(y+i\epsilon)\,-\,
\log(1-y+i\epsilon)\,-\pi i\bigr]
$$
which is chosen in such a way that
$\xi(-\infty)=0$, the solution of eq.(\ref{Atre}) is
\begin{eqnarray}
\label{solut}
\log \Phi_{01}\,&=&\,p\xi \,+\,
2\theta(-q)\,+\,2\theta(q-1)\,+
\,\frac 1{\pi i}\,\bigl[
\log(q-i\xi-i\epsilon)\,-
\,\log(q-i\xi+i\epsilon)\bigr]\,\nonumber\\
&+&\,\frac 1{\pi i}\,\bigl[
\log(1-q+i\xi-i\epsilon)\,-
\,\log(1-q+i\xi+i\epsilon)\bigr].
\end{eqnarray}

The $P\exp$ is actually needed only when the
upper limit is $y=\infty$, namely $\xi=-2\pi i a$, and,
whenever $a$ has an imaginary part, the $i\epsilon$ term in the logarithms
can be omitted.

In this case we obtain
\begin{equation}
\label{Aquattro}
\log \Phi_{01}(-2\pi i a,q)\,=-\,2\pi i a p.
\end{equation}
which reproduces the expression in eq.\,(\ref{duedue}). 

The first and the third pieces of $F$ in eq.\,(\ref{sum})
result in the same type of equations:
$$
\frac{\partial \Phi_{-\infty 0}}{\partial y}\,-\,
\varphi(y)\,\frac 1i\,\frac{\partial \Phi_{-\infty 0}}
{\partial q}\,=\,\varphi(y)\bigl[p\,+\,2i
\delta(q)\bigr]\,\Phi_{-\infty 0}
$$
and
$$
\frac{\partial \Phi_{1\infty}}{\partial y}\,-\,
\varphi(y)\,\frac 1i\,\frac{\partial \Phi_{1\infty}}
{\partial q}\,=\,\varphi(y)\bigl[p\,-\,2i
\delta(q-1)\bigr]\,\Phi_{1\infty},
$$
with the functions
$$
\varphi(y)\,=\,a\,\frac 1{y-i\epsilon}
\frac 1{1-y+i\epsilon}
$$
in the equation for $\Phi_{-\infty 0}$ and
$$
\varphi(y)\,=\,a\,\frac 1{y+i\epsilon}
\frac 1{1-y-i\epsilon}
$$
in the equation for $\Phi_{1\infty}$. The
solutions to these equations are
expressed by formulae analogous to (\ref{solut}) through
the variable $\xi$:
$$
\xi(y)\,=\,a\bigl[\log(y-i\epsilon)\,-\,
\log(1-y+i\epsilon)\,+\pi i\bigr]
$$
for $\Phi_{-\infty 0}$ and
$$
\xi(y)\,=\,a\bigl[\log(y+i\epsilon)\,-\,
\log(1-y-i\epsilon)\,-\pi i\bigr]
$$
for $\Phi_{1\infty}$. 

According to these definitions,
$\xi(-\infty)=0$ but, owing to the signs of $i\epsilon$,
$\xi(+\infty)=0$ too. Then $\log \Phi_{-\infty 0}=0$ and
$\log \Phi_{1\infty}=0$.

\end{document}